\documentclass[conference]{IEEEtran}
\usepackage{listings}
\usepackage{booktabs}
\usepackage{tcolorbox}
\IEEEoverridecommandlockouts
\usepackage{cite}
\usepackage{amsmath,amssymb,amsfonts}
\usepackage{algorithmic}
\usepackage{graphicx}
\usepackage{textcomp}
\usepackage{xcolor}

\usepackage[normalem]{ulem}

\usepackage[a4paper, total={184mm,239mm}]{geometry}

\usepackage{amsthm}
\newcommand{\figname}{{Fig.}}

\begin{document}

\title{Classification-Based Automatic HDL Code Generation Using LLMs\\
{}
}

\author{
\IEEEauthorblockN{Wenhao Sun$^1$,
Bing Li$^2$,
Grace Li Zhang$^3$,
Xunzhao Yin$^4$,
Cheng Zhuo$^4$,
Ulf Schlichtmann$^1$}
\IEEEauthorblockA{$^1$Technical University of Munich (TUM), 
$^2$ University of Siegen
$^3$TU Darmstadt, $^4$Zhejiang University
}
\IEEEauthorblockA{Email: \{wenhao.sun, ulf.schlichtmann\}@tum.de, bing.li@uni-siegen.de,  grace.zhang@tu-darmstadt.de,\{xzyin1, czhuo\}@zju.edu.cn}
}

\maketitle

\label{sec:abstract}

\begin{abstract}
While large language models (LLMs) have demonstrated the ability to generate hardware description language (HDL) code for digital circuits, they still suffer from the hallucination problem, which leads to the generation of incorrect HDL code or misunderstanding of specifications. In this work, we introduce a human-expert-inspired method to mitigate the hallucination of LLMs and improve the performance in HDL code generation.
We first let LLMs classify the type of the circuit based on the specifications. Then, according to the type of the circuit, we split the tasks into several sub-procedures, including information extraction and human-like design flow using Electronic Design Automation (EDA) tools. Besides, we also use a search method to mitigate the variation in code generation. Experimental results show that our method can significantly improve the functional correctness of the generated Verilog and reduce the hallucination of LLMs.
\end{abstract}
\section{Introduction}\label{sec:introduction}

As Moore's Law slowing down, there is an increasing demand for customized VLSI design. One of the key steps in the hardware design process is writing the hardware description language (HDL) code. However, HDL programming is time-consuming and labor-intensive. Therefore, the automatic HDL code generation from the specifications in natural language has attracted much attention in recent years. 

Among the automatic HDL code generation solutions, large language models (LLMs) code generation is one of the most promising solutions. LLMs have achieved remarkable success in various fields, such as machine translation \cite{translation} and robot trajectory planning \cite{robot}. In software design, LLMs have also demonstrated the ability to generate the code for various programming languages \cite{codegen}. In the field of hardware design, generative models have been employed to create designs \cite{survey}. Additionally, researchers also have recognized the potential of LLMs in generating HDL codes \cite{benchmark}.

\begin{figure}[ht!]
    \centering
    \includegraphics[width=0.48\textwidth]{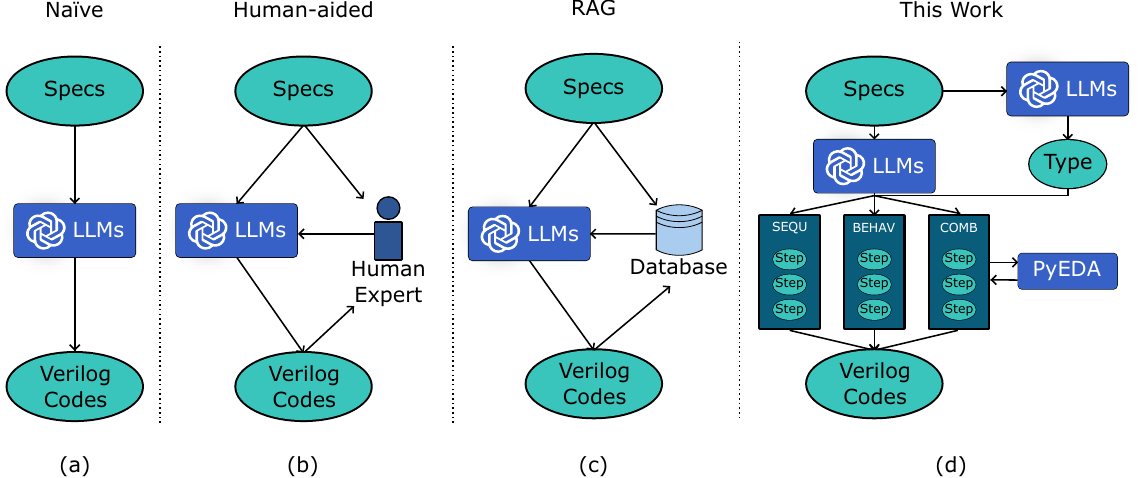}
    \caption{Comparison of the training-free methods. The naïve method suffers from the hallucination problem. The human-aided method requires human labor and is inefficient. The RAG method relies on the quality of the database which may be expensive to be built. By introducing the human-expert-inspired procedure, the proposed method can mitigate the hallucination of LLMs without human labor or database.
    }
    \label{full}
\end{figure}

Despite the advance of LLMs,
there are still many challenges that hinder the application of LLMs in HDL code generation. One of the challenges is the hallucination problem. The hallucination problem refers to the generation of incorrect HDL code or misunderstanding of the specifications. The naïve way as illustrated in Fig.~\ref{full}(a) to apply LLMs to generate HDL codes in one iteration usually leads to many errors.
The hallucination problem is caused by the lack of reasoning ability and training in the HDL domain, since the amount of datasets for HDL codes is limited. 
To solve this problem, a human expert can be involved 
 as a supervisor in the workflow of LLMs, such as Chip-chat \cite{humanaid1}, as illustrated in Fig.~\ref{full}(b), where human expert monitors the code generate process and provides to LLMs to correct the code. However, involving human experts increases the cost of this automatic design flow, and the effieceny of this work flow may also compromise the efficiency of the human expert.

ChipNeMo \cite{trained2} and \cite{trained1,trained3,trained4,trained5} aim to enhance the quality of generated HDL codes by fine-tuning the LLMs with augmented datasets. Therefore, the LLMs can learn more about the HDL design, which helps them to generate the correct HDL codes. However, this method depends on the quality of the augmented datasets. Besides, since fine-tuning an LLM needs a large amount of computation resources, the fine-tuned models are relatively small compared with commercial models such as GPT4 \cite{gpt4}. Accordingly, the reasoning ability of the fine-tuned LLMs may be limited. 

The other method to enhance the quality of automatically genearted HDL codes is to leverage the in-context learning ability \cite{incontext} of the LLMs. The in-context learning ability allows the LLMs to learn how to generate new patterns from the examples in the context. One of the applications is retrieval augmented generation (RAG) \cite{rag}, as illustrated in Fig.~\ref{full}(c), which retrieves the examples from a database to guide the code generation using LLMs. Previous work such as GPT4AIGChip \cite{raghdl1, raghdl2, raghdl3} uses RAG either to retrieve the examples similar to the specifications to help the generation or to retrieve the examples for error correction. However, these methods need a database with a large number of examples, which may need human experts to build. AutoChip \cite{autochip} uses the feedback from testbenches as a replacement for the database. Therefore, it can avoid the database building process. However, the testbenches do not always have feedbacks. In some scenarios, the final testbench can only show the pass rate of the test samples, which is not sufficient for the LLMs to correct the errors in code generation.

To mitigate the hallucination of LLMs without fine-tuning, human labor, database, or feedback from testbenches, we introduce a human-expert-inspired method to improve the performance in HDL code generation. 
As illustrated in Fig.~\ref{full}(d), similar to a human expert
we first let LLMs classify the type of circuit based on the specifications to decide the design procedures. Then, according to the type of the circuit, either the procedure for combinational logic (COMB) or the procedure for sequential logic (SEQU) is executed. In both procedures, LLMs extract the information from the specifications and generate an information list. The information list is further transited into a format that can be processed by Electronic Design Automation (EDA) tools or with
a divide-and-conquer strategy. Moreover, we also use a search method to distribute the test budgets and overcome the output variations of the LLMs. The contributions of this work are listed as follows:

\begin{itemize} 
    \item
    To mitigate the hallucination of LLMs, a human-expert-inspired procedure is introduced. Unlike the naïvely one-time HDL code generation, we split the tasks into several sub-procedures, similar to the procedure that a human expert uses to design the circuits. The sub-procedures are designed to match the ability of the LLMs and the characteristics of the tasks. 

    \item 
    We use a search method to distribute the test budgets to the \textit{type-specific} procedures (COMB and SEQU) and \textit{general} procedure (BEHAV). This strategy improves the efficiency of the code generation procedures by selecting the most promising information list and the distribution between the \textit{type-specific} and \textit{general} procedures.

    \item
    Experimental results demonstrate that our method can significantly improve the functional correctness of the generated HDL codes. The proposed method outperforms the baseline in the VerilogEval-human dataset\cite{veval} with 4.7\%, 11.0\%, and 14.7\% in Pass@1, Pass@5, and Pass@10, respectively. In VerilogEval-machine dataset, the proposed method also improves more than 5\% pass rate in Pass@5 and Pass@10. 
\end{itemize}

\section{Motivation}\label{sec:motivation}
To improve the performance of Large Language Models (LLMs) in HDL code generation in a training-free manner, the key is to use the in-context learning and reasoning ability of the LLMs. However, RAG-based methods heavily rely on the quality of the databases, which are also expensive to build. On the other hand, taking feedback from testbenches as a replacement for the database is not always feasible. If the coverage of testbenches is not sufficient, the feedback may drive the LLMs to generate the codes that can only pass the testbenches, which may not be the exact design consistent with the specifications. If the coverage of testbenches is sufficient, there may be too much feedback, which may then overwhelm the LLMs. Without the help of external information, our solution aims to improve the reasoning ability of the LLMs in HDL generation by reducing the design space of the HDL codes using LLM-based classification.

Recent research \cite{hop} on the reasoning ability of LLMs reveals that 
LLMs suffer from multi-hop reasoning. For example, if we want the LLMs to take a value $a$ from an array $A[10]$ and then use $a$ as an index to access another array $B[10]$, the LLMs may successfully provide the correct answer. However, if we further ask the LLMs to use the value from $B$ as an index to access another array $C[10]$, the LLMs are likely to fail. Accordingly, the reasoning ability of state-of-the-art LLMs is usually limited to one-hop reasoning.

According to the discussion above, we should shorten the reasoning hops in HDL generation using LLMs. In other words, we should extract as many conditional decisions as possible in this procedure. Accordingly, we split the task of HDL generation in conditional several sub-tasks, whose numbers of reasoning hops are smaller than the direct code generation from specifications and thus do not exceed the reasoning ability of the LLMs. This strategy is similar to
Chain-of-Thought (CoT) \cite{cot} in improving the reasoning ability of the LLMs using the self-planning ability of the LLMs themselves. 
This method performs well in scenarios such as \textit{general-purpose} QA. 
However, the CoT method is not suitable for the HDL code generation tasks,
%
because the training datasets of HDL design are relatively small and usually lack comprehensive information about the low-level details of circuit design knowledge, e.g., combinational logic and sequential circuits. Consequently, LLMs struggle with self-planning during the HDL design process and cannot directly use the internal knowledge of the design methodology without explicit guidance.
Referring to the design flow of human experts, we aim to design a method to guide the LLMs in generating the HDL codes in a human-like procedure, while keeping the sub-tasks within the reasoning ability of the LLMs.

\section{Proposed Approach}\label{sec:method}

\begin{figure}
    \centering
    \includegraphics[width=0.48\textwidth]{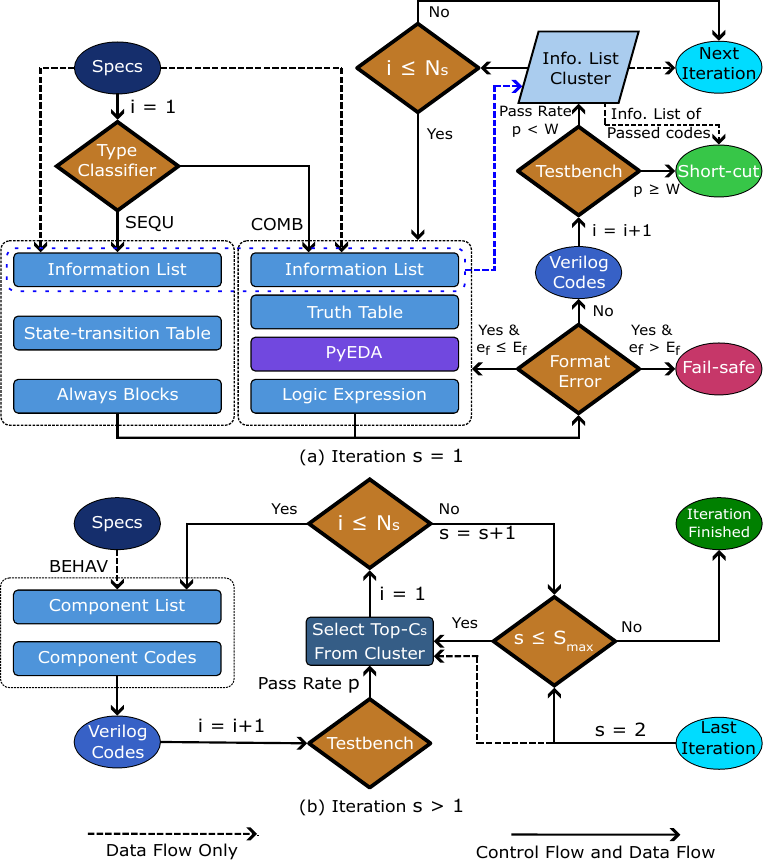}
    \caption{The workflow of the proposed method. The first iteration is illustrated in (a), where the specifications are classified. Depending on the circuit type, either the SEQU or COMB procedure is executed and verified with testbench. The further iterations are illustrated in (b) and start at \textit{Last Iteration}, where the information list is reused and the BEHAV procedure is executed.
    }
    \label{chain}
\end{figure}

In this section, we introduce the proposed method in detail. 
%
%
%
%
Fig.~\ref{chain} shows the workflow of our approach. Similar to human experts, who leverage their experience and knowledge to determine the appropriate design types for a specific task, we start by classifying the type of the circuit based on the specification. 
The typical types include combinational and sequential circuits. However, for a specification, if the code generation based on the types cannot generate the correct codes, it will be considered as general type and processed by a general procedure. With this classification, the required depth of reasoning by LLMs can thus be reduced, so that the hallucination of LLMs in subsequent steps can be mitigated due to more specific available information.

After circuit type classification, LLMs can thus generate HDL design according to these types. SEQU is the procedure for sequential logic, and COMB is for combinational logic. There is also a general procedure, BEHAV, to handle the tasks that are hard to process with SEQU or COMB. To take advantage of the available type information, we also transform the original design specification into a specific information list with respect to each circuit. This explicit specification is a further step in reducing the depth of reasoning and enhancing the quality of generated Verilog codes. 

%
The code generation  above, illustrated within the dashed rectangles in Fig.~\ref{chain} (a), is executed in a loop to produce $N_s$ code candidates, where $N_s$ is a configurable parameter of the proposed method.
The generated code candidates are evaluated using a testbench to assess their functional correctness, which is measured by the pass rate $p = \frac{m}{n}$, where $m$ is the number of passed samples and $n$ is the number of total samples in the testbench. Here, the sample in testbench means a stimulus/response pair.
In the first iteration, this evaluation determines whether the code generation procedure should iterate normally or switch to \textit{Short-cut} mode.
If all the generated codes cannot pass the testbench, the code generation procedure is repeated to further iterations. 
In the new iteration, the procedure will be switched to BEHAV. The iterations end when the number of iteration $s$ reaches the max iteration $S_{max}$.

In the code generation iterations shown in Fig.~\ref{chain}, we assume that the total number of testbench executions to verify the generated codes is a fixed number, denoted as $N_{total} = \sum{N_s}$. This assumption is based on the practical consideration that verification is time-consuming and expensive. To achieve good efficiency, we implement a search method to distribute these test budgets among the generated codes. Since the LLMs have variations in code generation, it is necessary to find and select a better intermediate result, thus enhancing the possibility of generating the correct codes. Since information lists are the foundation of the subsequence steps, the search method mitigates the impact of variation via finding the most promising information list from the candidates generated in the first iteration. Then, unlike the first iteration, the BEHAV procedure is used to generate the codes from the selected information lists.
The details of the workflow in  \figname~\ref{chain} are described as follows.


\subsection{Circuit Classification}

For a given specification, the first task is to classify the circuit type using LLMs, as \textit{Type Classifier} shown in \figname~\ref{chain}. This classification only needs to be performed once, and the circuit type will be shared in all iterations.
%
%
A naïve approach to classification is to ask LLMs to 
directly determine the type of the circuit based on the specifications.
However, LLMs may have hallucinations during this process, because they are not specifically trained for this kind of tasks. In addition, the necessary information for classification may be obscured by the semantics or the style of the language description of the specifications. For instance, if the specification only requires a combinational logic that functions within a sequential circuit, LLMs may incorrectly classify the circuit as a sequential logic. 

To mitigate this hallucination, we first ask LLMs to generate Verilog codes for the target circuit using the given design specification. Since LLMs may have the ability to capture the structure of the circuit, the resulting Verilog codes can be used to deduce the target circuit types, although the circuit may not be functionally correct.




\subsection{Information List Extraction}

After circuit type classification, the LLMs proceed to extract more specific information from the design specifications, As illustrated in \figname~\ref{chain} \textit{Information List}. The purpose of the information extraction is to transform the implicit information contained within the specifications into explicit information about the relationship between the module inputs and outputs. This transformation also helps eliminate redundant information, such as descriptions of the application scenarios, which could lead to hallucinations in the later steps. 

The extraction process varies between circuit types. For combinational logic, the information list forms the truth table as depicted in \figname~\ref{infolistcombapp} in the Appendix, while for sequential logic, besides the relationship between the inputs and outputs, it is crucial to include timing descriptions in the information list, as shown in \figname~\ref{infolistsequapp} in the Appendix.

Observing the flow in Fig.~\ref{chain}, all the information lists are generated in the first iteration and stored in the information list cluster with their testbench pass rate $p$, as shown in \textit{Info. List Cluster}. Apparently, all three procedures either directly rely on the newly generated information list or rely on the information list selected from the last iteration. If there is missing  or incorrect information in the list, the error will be propagated to subsequent steps, such as truth table and state-transition table generation, potentially impacting the final generated codes. On the other hand, an information list of good quality benefits the subsequent steps by providing comprehensive information and emphasizing the crucial information, thereby improving the overall effectiveness of the code generation process. 

\subsection{\textit{Type-specific} Procedures for Circuit Geneartion}

After information list extraction in the first iteration, the next step is to instruct LLMs to convert the extracted information list into a standardized format, as depicted in the dashed rectangles in Fig.~\ref{chain} (a). These formats naturally break down the tasks into manageable sub-tasks that fall within the reasoning capabilities of the LLMs. Additionally, the formatted data, such as the truth table, can be captured by scripts and further processed by EDA tools.

For the sequential logic procedure, SEQU, this format is the state-transition table, which is derived from the information list by the reasoning ability and internal knowledge of LLMs, as shown in the example in Fig.~\ref{sequproc} in Appendix. For the combinational logic procedure, COMB, it is the JSON formatted truth table, as shown in the example in Fig.~\ref{combproc} in Appendix. 

In the next step, depending on the type of the circuit, the LLMs generate the codes by following the design paradigms. For sequential logic, as shown in Fig.~\ref{chain}(a) SEQU, we employ the three-always-block method to complete the tasks. To ensure the sub-tasks remain within the reason capabilities of the LLMs, we instruct the LLMs to complete the \textit{always block}s sequentially. Before generating the code of an \textit{always block}, we allow the LLMs to produce a description of this block, providing space for reasoning and potentially minimizing hallucination. Subsequently, the LLMs merge the generated \textit{always block}s into a complete module, as shown in Fig.~\ref{merge1} and Fig.~\ref{merge2} in Appendix.

For combinational logic, the standard format is the truth table formatted in JSON, which, unlike the truth table in information list, can be easily captured and processed by external scripts, as shown in Fig.~\ref{chain} (a) COMB. For LLMs, performing calculations is a complex task, as they tend to provide answers that resemble the correct response rather than doing detailed calculation steps like human. These answers are usually incorrect. Therefore, LLMs have yet to obtain the ability to simplify the truth table correctly. However, since the truth table is already formatted into JSON, we utilize PyEDA to simplify it into a sum-of-products (SOP) expression. This approach is faster and more accurate than relying solely on the LLMs. Subsequently, the LLMs only need to generate the Verilog codes based on the SOP expression.

To mitigate the variation of LLMs, the SEQU or COMB procedure is performed $N_i$ times and generates $N_i$ code samples. These code samples are tested by the testbench, and their quality are measured. The testbench pass rate $p$ not only indicates the code quality but also is a sign of the information list quality. Thus, the testbench pass rate works as a score and is marked on the corresponding information list. Then, it is stored in the information list cluster for the next iteration.

\subsection{\textit{General} Procedure for Circuit Geneartion}
After the first iteration, tasks without candidates passing the testbench enter the next iteration. For tasks that are challenging to be converted into a standard format, which is indicated by the low pass rate in the first iteration, the behavior design diagram, denoted as BEHAV, is utilized, as shown in Fig.~\ref{chain}(b). In the BEHAV procedure, the information list generated in the first iteration is selected and reused. We select the top-$C_s$, where $C_s$ is a pre-defined value, information lists from the cluster according to their pass rates, like beam search \cite{beamsearch}. Similar to the procedure for sequential logic that emphasizes the \textit{always blocks}, the behavior design diagram focuses on the \textit{components}, such as a code block of a \textit{for} loop to describe an adder with high bit-width, where the truth table is hard to be enumerated or it contains both combinational and sequential logic circuits. We instruct LLMs to divide the task into several components, each associated with different sub-tasks. First, these components are organized into one component list with descriptions. Then, the LLMs generate the codes for a component at a time. Finally, the LLMs integrate these components into a complete module.

In each iteration, the BEHAV procedure is executed $N_s$ times and produces $N_s$ code samples. Afterward, it is tested by the testbench. Like the \textit{Type-specific} procedures, the information lists should also be stored in the cluster and marked with the scores. Since in former iterations, the information lists are already marked with pass rates, to fully use this information, we use the average pass rate of the current iteration and former iterations, denoted as $\frac{\sum p_s}{s_c}$, where $p_s$ is the pass rate in iteration $s$, and $s_c$ is the index of the current iteration.

  


\begin{table*}[!ht]
  \centering
    \caption{Comparison of Pass@$k$ for the proposed method and baseline.}
  \begin{tabular}{c c c c c c c c c c c c c c c}
  \hline
  \toprule
   Pass@$k$ & & \multicolumn{6}{c}{VerilogEval-human} & & \multicolumn{6}{c}{VerilogEval-machine} \\
\cline {1-1} \cline {3-8} \cline {10-15}
\addlinespace[2pt]
    \% & & Baseline & FULL & COMB & SEQU & BEHAV & SYNTAX & & Baseline & FULL & COMB & SEQU & BEHAV & SYNTAX\\
  \hline
  \addlinespace[2pt]
  Pass@1 & & 42.0 & 46.7 & 44.6 & 43.7 & 42.5 & 86.9 & & 63.8 & 65.2 & 63.8 & 65.1 & 63.9 & 92.9\\ 
  Pass@5 & & 58.6 & 69.6 & 62.2 & 63.9 & 60.7 & 98.8 & & 76.3 & 81.4 & 76.3 & 80.7 & 77.0 & 98.9\\ 
  Pass@10 & & 62.2 & 76.9 & 66.0 & 69.2 & 66.0 & 99.4 & & 78.3 & 85.3 & 78.3 & 83.9 & 79.7 & 99.3\\ 

  \hline \toprule\\
  
  \end{tabular}

  \label{main}
\end{table*}

\subsection{\textit{Fail-safe} and \textit{Short-cut} Mode}\label{sec:specialmode}
To handle special situations in the first iteration, two strategies, \textit{Fail-safe} and \textit{Short-cut}, are incorporated, as depicted in Fig.~\ref{chain}(a) \textit{Fall-safe} and \textit{Short-cut}. 

Since the LLMs have variation in the generated answers, sometimes the script in the later step cannot capture the intermediate results from the former step. We define this error as \textit{Format Error}. We allocate each task a total of $E_f$ chance for the format error retry, where intermediate outputs of LLMs are deprecated and the current procedure is executed again. The number of current \textit{Format Error}s is denoted as $e_f$. If $e_f > E_f$, we conclude that the \textit{type-specific} procedures are unsuitable for the task, indicating that LLMs cannot answer or generate an incorrect answer. In this case, the iteration switches to \textit{general} procedure and enters \textit{Fail-safe} mode. In \textit{Fail-safe} mode, we allocate all remaining test budgets to the \textit{general} procedure and regenerate the information lists before each execution of the \textit{general} procedure.

Conversely, if we identify a promising information list that allows the \textit{type-specific} procedures to nearly pass the testbench, we should allocate all remaining test budgets to this information list and keep using the previous procedure. We denote this mode as \textit{Short-cut} mode. We establish a threshold $W$ for the testbench sample pass rate $p$. If the testbench sample pass rate $p$ exceeds $W$, we consider the information list to be the seed player, and the corresponding procedure is sufficiently robust for the task. In this scenario, the search process switches to \textit{Short-cut} mode, where all remaining test budgets are assigned to the current procedure, and the current information list is also kept.
 
\section{Experimental Results}\label{sec:results}

To demonstrate the performance of the proposed method, we conducted experiments on the VerilogEval \cite{veval} dataset, which contains 299 tasks with specifications, testbench, and ground truth Verilog codes. In this dataset, the tasks are divided into two categories: VerilogEval-human and VerilogEval-machine. The specifications of VerilogEval-human tasks are written by human experts, while the VerilogEval-machine tasks are generated by gpt-3.5-turbo\cite{gpt35} according to the ground truth codes.

\begin{figure}[t]
  \centering
  \includegraphics[width=0.48\textwidth]{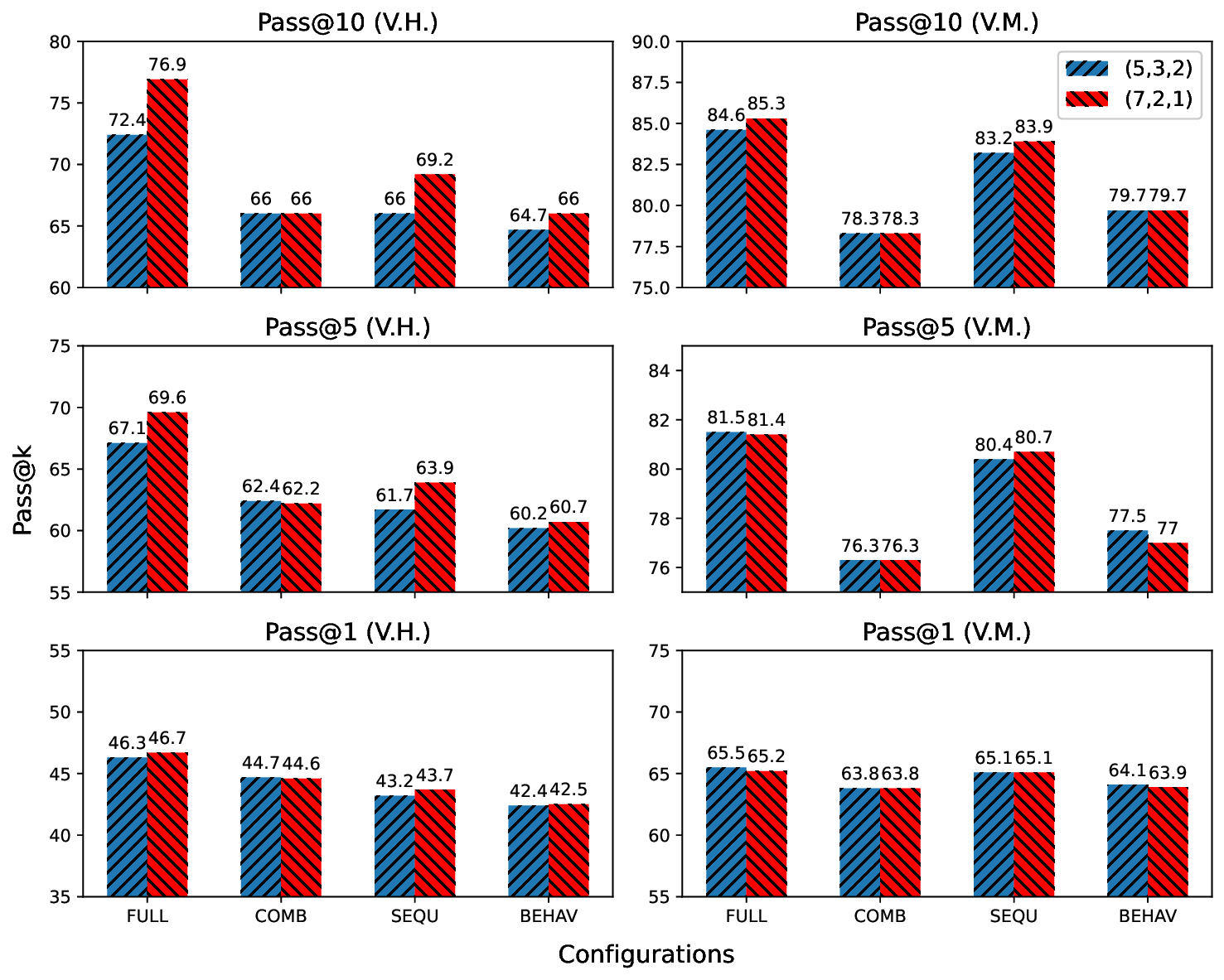}
  \caption{Comparison of the Pass@k between two configurations, (5,3,2) and (7,2,1), where V.H. represents VerilogEval-human, and V.M. represents VerilogEval-machine. The total results of all the procedures are shown as FULL, and the individual results of the COMB, SEQU and BEHAV procedures are shown as COMB, SEQU and BEHAV, respectively.}
  \label{configcompare}
\end{figure}

For the evaluation, we adopted GPT4 \cite{gpt4} in the proposed framework through OpenAI APIs. 
Despite the newer GPT4s released, such as gpt-4-0125-preview, we chose gpt-4-0613 as our base model, because it was released before the VerilogEval dataset was published. Therefore, the training sets of this GPT4 version do not contain the VerilogEval dataset. The temperature of the LLM was set to $0.5$ for all the experiments. The maximum context length was set to $4096$. For simulation, we used Icarus Verilog (iverilog) \cite{iverilog} to run the testbench and the generated codes.

We used the Pass@$k$ \cite{passat} as the metric to evaluate the functional correctness of the generated codes. In Pass@$k$, $k$ code samples are tested and we evaluate the number of the samples passed the testbench. It is defined as
\begin{equation}
    Pass@ k:=\underset{Problems }{\mathbb{E}}\left[1-\frac{\binom{n-c}{k}}{\binom{n}{k}}\right]
\end{equation}
where $n$ is the sample number per task, and $c$ is the number of samples that passed the testbench.
Here, we consider that every execution of testbench is counted as a code sample tested in Pass@$k$. We set $n$ to 10 and $k$ to $1$, $5$, and $10$ in the experiments, respectively. The maximum format retry $E_f$ was set to $10$ for both the proposed method and the baseline, in which the LLM directly generates the codes from specifications. For the search parameters, we set the max iteration $S_{max}$ to 3. Thus the configuration of the number of generated code samples can be expressed as $(N_1,N_2,N_3)$, which is set to $(7, 2, 1)$. The Top-$C_s$ is set as $C_2=N_2$ and $C_3=1$, which means in the second iteration, the two code samples are generated from top-$2$ information lists, and in the last iteration, all the rest ot the test budgets are given to the top-$1$ information list. The \textit{Short-cut} threshold \textit{W} is set to 0.95. To reduce the token cost, we first conducted the baseline experiments. Then, we selected the difficult tasks, for which the baseline method cannot generate valid codes to pass the test after $10$ retries, to be processed by the proposed framework. 

In Table \ref{main}, we compare the baseline and the proposed method. In this table, Baseline is the direct generation using GPT-4. FULL and SYNTAX are the results of the functional correctness and syntax correctness of the proposed method, respectively. The contributions of the COMB, SEQU and BEHAV procedures are individually shown as COMB, SEQU and BEHAV in the columns, respectively. 
For VerilogEval-human, the proposed method outperforms the baseline with 4.7\%, 11.0\%, and 14.7\% in Pass@1, Pass@5, and Pass@10, respectively. In VerilogEval-machine, the proposed method also improves the pass rate in all Pass@$k$ settings, where the improvements are over 5\% in Pass@5 and Pass@10. 

In Table \ref{main}, it is worth noting that all three procedures, COMB, SEQU, and BEHAV, contributed to the improvements in the VerilogEval-human dataset. However, the COMB procedure had no improvement in the VerilogEval-machine dataset. This is because the specifications in VerilogEval-machine dataset are generated by summarizing the ground truth codes with LLMs, which means the logical relationship between the inputs and outputs is already shown in the specifications. Hence, the LLMs can bypass the calculation and reasoning procedure in solving combinational logic tasks, such as simplifying the truth table.

\begin{figure}
  \centering
  \includegraphics[width=0.48\textwidth]{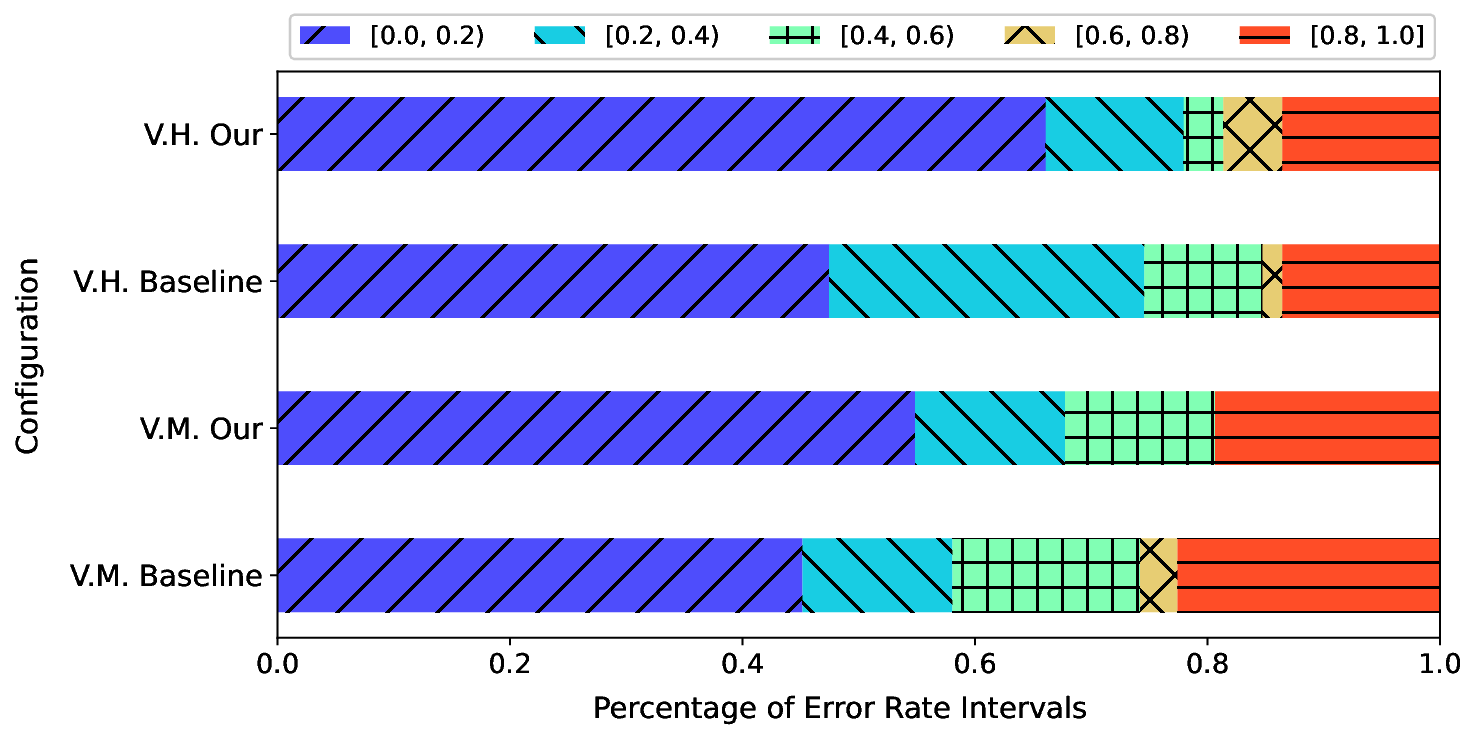}
  \caption{Comparison of the task error rate distributions in difficult tasks, where for each task, the error rate is from the code sample with the best performance in Pass@$10$. The x-axis represents the percentage of the error rate intervals, and the y-axis represents the configurations. V.H. represents VerilogEval-human, and V.M. represents VerilogEval-machine. }
  \label{distribution}
\end{figure}

To demonstrate the influence of the search parameter, the results of the search configuration $(5,3,2)$ and $(7,2,1)$ are compared in Fig.~\ref{configcompare}. We observe that the configuration $(7,2,1)$ is better than $(5,3,2)$ in most cases except for Pass@1 and Pass@5 in VerilogEval-machine. Overall, the configuration $(7,2,1)$ is better than $(5,3,2)$ in the VerilogEval dataset. The potential reason is that configuration $(7,2,1)$ has more information list candidates, contributing to a higher possibility of finding a high-quality information list.

\begin{table}[!t]
  \centering
    \caption{Comparison of Pass@$10$ For AutoChip and the proposed Method in Ten Tasks of VerilogEval-human}
  \begin{tabular}{c c c}
  \hline
  \toprule
   \% & AutoChip & Our method \\
  \toprule
  \addlinespace[2pt]
  Sequential logic & 40.0 & 50.0 \\
  Combinational logic & 0.0 & 70.0 \\
  \hline \toprule\\
  
  \end{tabular}

  \label{autochip}
\end{table}

\begin{table}[!t]
  \centering
    \caption{Comparison of Type Classification Accuracy with and without Codes.}
  \begin{tabular}{c c c}
  \hline
  \toprule
   \% & w/ code & w/o code \\
  \toprule
  \addlinespace[2pt]
  VerilogEval-human & 98.3 & 94.9 \\
  VerilogEval-machine & 100.0 & 100.0 \\
  \hline \toprule\\
  
  \end{tabular}

  \label{type}
\end{table}

The task error rate distributions of the baseline and the proposed method in the hard tasks of VerilogEval are illustrated in Fig.~\ref{distribution}, where the error rate for each task is from the best code sample in Pass@10. We split the error rate distributions into five intervals with step 0.2. The proportions of the intervals are shown on the x-axis. 
It can be observed that the proposed method not only improves the Pass@k but also reduces the code sample error rates. However, the number of the hardest tasks, whose error rates are higher than 0.8, as shown in the rightmost of the bars, did not reduce in VerilogEval-human and is only slightly lower in VerilogEval-machine. This indicates that there remains room for improvement in the proposed method.

\begin{figure}[t]
 \small 
\rule{\linewidth}{0.4pt}
\vspace{0mm} 
\emph{... \\
  6. q: This is a 64-bit register which holds the contents of the shift register. It gets updated with the new value either when the shift operation takes place (if ena is high) or when the new data is loaded into the shift register (if load is high). The updating of this register takes place on the positive edge of the clk signal. \\
  ... \\
}
\rule{\linewidth}{0.4pt}
\emph{... \\
  6. q: This is the 64-bit output of the shift register. It represents the current contents of the shift register. \\
  ... \\
}
\rule{\linewidth}{0.4pt}

  \caption{Comparison of snippets of the information lists between two attempts of generation in a sequential logic task. The upper section is from the attempt where the codes passed the testbench, while the lower section is from the attempt where the codes failed during simulation.}
  \label{infolistsequ}
\end{figure}

\begin{figure}
\small
\rule{\linewidth}{0.4pt}
\vspace{0mm} 
\emph{    ... \\
    14. When a=0, b=0, c=1, d=0, the output is 1.\\
    15. When a=1, b=0, c=1, d=0, the output is 1.\\
    ... \\
    }
\rule{\linewidth}{0.4pt}
\emph{    ... \\
    6. When a=0, b=0, c=1, d=0, the output is 0.\\
    7. When a=1, b=0, c=1, d=0, the output can be 0 or 1 (don't care). \\
    ... \\
    }
\rule{\linewidth}{0.4pt}

  \caption{Comparison of snippets of information lists between two attempts of generation in a combinational logic tasks. The upper section is from the attempt where the codes passed the testbench, while the lower section is from the attempt where the codes failed during simulation.}
  \label{infolistcomb}
\end{figure}

In Table \ref{autochip}, the performance of AutoChip \cite{autochip} and the proposed method are compared. For both sequential and combinational logic, we randomly chose ten tasks from the difficult tasks of VerilogEval-human dataset. Here, to evaluate the performance of the methods under the same test standard, we followed the test standard in Pass@$10$, where each execution of the testbench is counted as a test, regardless of whether it happened inside an iteration or after a complete attempt of the method. For AutoChip, the iteration of the feedback, $n$, was set to $10$. The results show that the proposed method performs slightly better on sequential logic and significantly better on combinational logic. The reason is that AutoChip heavily depends on the feedback of the testbench. However, the testbench in VerilogEval only provides limited hints for sequential logic and almost no hints for combinational logic. On the contrary, the proposed method does not need the feedback of the testbench and effectively compensates for the shortcomings of LLMs in handling combinational logic.

Table \ref{type} presents a comparison of type classification accuracy between methods with and without the usage of the generated codes. In the VerilogEval-machine dataset, both methods can classify the type of the circuit correctly. However, in the VerilogEval-human dataset, the method using codes generated by LLMs has a 1.7\% error rate, while the naïve method has a 5.1\% error rate. Analysis of the error cases reveals that the errors are due to misleading specifications, which require combinational logic parts from a sequential logic circuit. The naïve method focuses more on the semantics of the specifications, while the method using codes generated by LLMs can leverage the internal knowledge of the Verilog design and avoid some errors.

The comparison of the information lists between two examples whose codes passed and failed in the simulation is shown in Fig.~\ref{infolistsequ} and Fig.~\ref{infolistcomb}. In the case of the sequential logic task, as shown in Fig.~\ref{infolistsequ}, the information list of the failed codes contains fewer descriptions compared to the list of the codes that passed in the simulation. Because of the variation of LLMs, some information lists might be generated with information loss, leading to the generation of failed codes. While detailed descriptions might be considered redundant by human experts, they help the LLMs to concentrate on the key information. In the examples of combinational logic task, as shown in Fig.~\ref{infolistcomb}, the information list, which failed in the simulation, has an incorrect output description. Both the outputs, (a=0, b=0, c=1, d=0) and (a=1, b=0, c=1, d=0), are incorrect, which is caused by the hallucination of the LLMs. The correct outputs should be the information list of the codes that passed the testbench.
\section{Conclusion}
\label{sec:conclusion}

In this work, we proposed a human-expert-inspired method to improve the performance in HDL code generation using LLMs. We first let LLMs classify the type of the circuit based on the specifications. Then, according to the type of the circuit, three different design procedures are used to mitigate the hallucination of LLMs. Besides, we use a search method to distribute the test budgets efficiently. The experimental results show that our method can significantly improve the functional correctness of the generated HDL.

\let\oldbibliography\thebibliography
\renewcommand{\thebibliography}[1]{%
\oldbibliography{#1}%
\fontsize{6.85pt}{6.85}\selectfont
\setlength{\itemsep}{0.1pt}%
}

\clearpage
\appendix
\section{Examples}
\label{sec:appendix}

Here are some examples of the procedures described in Section \ref{sec:method}.

\begin{figure}[h]
{
\rule{\linewidth}{0.4pt}
\textbf{Sequential Logic}\\
\fontsize{10}{12pt}\selectfont
Specification:
\linespread{2.0} \selectfont
}

{
    \fontsize{7pt}{5pt}\selectfont
    \begin{verbatim}
Build a decade counter that counts from 0 through 9,
inclusive, with a period of 10. The reset input is 
active high synchronous, and should reset the counter
to 0. We want to be able to pause the counter rather 
than always incrementing every clock cycle, so the 
"slowena" input if high indicates when the counter 
should increment. 
    \end{verbatim}
}
\fontsize{10}{12pt}\selectfont
Prompt: \textit{List the relationship between the inputs and outputs one by one with details as many as possible. Do not simplify! Keep everything as the same in the task. If it has reset and clk, list the timing requirements of reset and clk.\\}
LLM:\emph{
1. clk: This is the clock input to the module. The decade counter will increment based on this clock signal. The timing requirement for this signal would be that it must be periodic and consistent to ensure accurate counting.\\
...\\
4. q: This is the output of the module. It is a 4-bit signal that represents the current count of the decade counter. This signal will change at the rising edge of the clock if slowena is high and reset is low. If reset is high, q will be reset to 0 regardless of the state of slowena. \\
}
\rule{\linewidth}{0.4pt}
    \caption{Example of information list generated by LLMs for sequential logic task.}

    \label{infolistsequapp}
\end{figure}

\begin{figure}[!h]
{
\rule{\linewidth}{0.4pt}
\textbf{Combinational Logic}\\
\fontsize{10}{12pt}\selectfont
Specification:
\linespread{2.0} \selectfont
}

{
    \fontsize{7pt}{5pt}\selectfont
    \begin{verbatim}
Implement the circuit described by the Karnaugh map below.

//       ab
// cd   00 01 11 10
//  00 | 1 | 1 | 0 | 1 |
//  01 | 1 | 0 | 1 | 1 |
//  11 | 0 | 0 | 1 | 0 |
//  10 | 1 | 1 | 1 | 0 |
    \end{verbatim}
    }
\fontsize{10}{12pt}\selectfont
Prompt: \\
\textit{List the relationship between the inputs and outputs one by one with details as many as possible. Do not simplify! Keep everything as the same in the task. Do not write the codes! If it has reset and clk, list the timing requirements of reset and clk.\\}
LLM:\\
\emph{
Sure, here are the relationships between the inputs and output according to the Karnaugh map:\\
1. When a = 0, b = 0, c = 0 and d = 0, out = 1\\
...\\
16. When a = 1, b = 0, c = 1 and d = 0, out = 0\\
These are the complete relationships between the inputs and output as per the Karnaugh map, without any simplification. \\
}
\rule{\linewidth}{0.4pt}
    \caption{Example of information list generated by LLMs for combinational logic task.}
    \label{infolistcombapp}
\end{figure}

\begin{figure}[!h]
\rule{\linewidth}{0.4pt}
\textbf{Sequential Logic}\\
\fontsize{10}{12pt}\selectfont
Information List: ...\\
Prompt: \\
\textit{Give the state transition table with details as many as possible. Do not simplify! Keep everything as the same in the task. Do not give the verilog codes.}\\
LLM: \\
\textit{Sure, we can define a state machine with four states: IDLE, BYTE1, BYTE2, and BYTE3. Here's the state transition table:}\\
    \fontsize{7pt}{5pt}
    \begin{verbatim}
    | Current State | in[3] | reset | Next State | done |
    |---------------|-------|-------|------------|------|
    | IDLE          | 0     | 0     | IDLE       | 0    |
    | IDLE          | 0     | 1     | IDLE       | 0    |
    | IDLE          | 1     | 0     | BYTE1      | 0    |
    | IDLE          | 1     | 1     | IDLE       | 0    |
    | BYTE1         | X     | 0     | BYTE2      | 0    |
    | BYTE1         | X     | 1     | IDLE       | 0    |
    | BYTE2         | X     | 0     | BYTE3      | 0    |
    | BYTE2         | X     | 1     | IDLE       | 0    |
    | BYTE3         | 0     | 0     | IDLE       | 1    |
    | BYTE3         | 0     | 1     | IDLE       | 0    |
    | BYTE3         | 1     | 0     | BYTE1      | 1    |
    | BYTE3         | 1     | 1     | IDLE       | 0    |
    \end{verbatim}
    \fontsize{10pt}{12pt}\selectfont
\textit{In this table, 'X' denotes a don't care condition. The 'done' signal is asserted high (1) only after the third byte of each message is successfully received, and it's de-asserted (0) in the next cycle. If 'reset' is asserted high (1), the state machine returns to the IDLE state, regardless of the current state and the value of in[3].}
\rule{\linewidth}{0.4pt}

    \caption{Example of the state-transition table generated by LLMs for a sequential logic task, where the information list is omitted.}
    \label{sequproc}
\end{figure}

\begin{figure}[!h]
\rule{\linewidth}{0.4pt}
\textbf{Combinational Logic}\\
\fontsize{10}{12pt}\selectfont
Information List: ...\\
    Prompt: \\
    \textit{Give the truth table between the outputs and inputs in json. Do not simplify! Keep everything as the same in the task. Do not give the verilog codes. Use the json format like this example:}
    \fontsize{7pt}{5pt}
    \begin{verbatim}
    {
        "table": [
        [0,0,0],
        [0,1,1],
        [1,0,1],
        [1,1,1]
        ],
        "inputs": ["a[1]","a[2]"],
        "outputs": ["x"]
        "header_inputs": ["a[2]","a[1]"],
        "header_outputs": ["x"]
    }
    \end{verbatim}
\fontsize{10pt}{12pt}\selectfont
LLM: \\
\textit{Here is the truth table in json format}
\fontsize{7pt}{5pt}
    \begin{verbatim}
    {
        "table": [
            [0,0,0,0,0], [0,1,0,0,1],
            [1,0,0,0,1], [1,1,0,0,1],
            [0,1,0,1,0], [0,0,0,1,0],
            [1,0,0,1,1], [1,1,0,1,1],
            [0,1,1,1,0], [0,0,1,1,1],
            [1,0,1,1,1], [1,1,1,1,1],
            [0,1,1,0,0], [0,0,1,0,1],
            [1,0,1,0,1], [1,1,1,0,1]
        ],
        "inputs": ["a","b","c","d"],
        "outputs": ["out"],
        "header_inputs": ["a","b","c","d"],
        "header_outputs": ["out"]
    }
    \end{verbatim}
\rule{\linewidth}{0.4pt}
    \fontsize{10pt}{12pt}\selectfont
    \caption{Example of the LLMs generated JSON format truth table for a the combinational logic task. The prompt contains an example for LLMs to constrain the output format. The information list is omitted.}
    \label{combproc}
\end{figure}

\begin{figure}[!h]

\rule{\linewidth}{0.4pt}
\fontsize{10}{12pt}\selectfont
\textbf{Generated Always Blocks}\\

\textit{The three always blocks in Verilog for this task will be as follows:}

\textit{1. State Register: This block will be used to update the state of the shift register. This block will be sensitive to the clock signal and will update the state of the shift register on the rising edge of the clock if the enable signal is high. The new state of the shift register will be determined by the next state logic.}

{
    \fontsize{7pt}{5pt}\selectfont
    \begin{verbatim}
always @(posedge clk) begin
	if(enable) begin
		Q <= nextState;
	end
end
    \end{verbatim}
}
\fontsize{10}{12pt}\selectfont
\textit{2. Next State Logic: This block will determine the next state of the shift register based on the current state and the S input. The next state will be the current state shifted to the left by one bit, with the MSB being the S input. This block will be combinational and will be sensitive to the current state of the shift register and the S input.}

{
    \fontsize{7pt}{5pt}\selectfont
    \begin{verbatim}
always @(Q, S) begin
	nextState = {Q[6:0], S};
end
    \end{verbatim}
}
\textit{3. Output Logic: This block will determine the output Z based on the current state of the shift register and the inputs A, B, and C. The inputs A, B, and C will be used as select lines for the multiplexer. The output Z will be the selected bit from the shift register. This block will be combinational and will be sensitive to the current state of the shift register and the inputs A, B, and C.}

{
    \fontsize{7pt}{5pt}\selectfont
    \begin{verbatim}
always @(Q, A, B, C) begin
	case({A, B, C})
		3'b000: Z = Q[0];
		3'b001: Z = Q[1];
		3'b010: Z = Q[2];
		3'b011: Z = Q[3];
		3'b100: Z = Q[4];
		3'b101: Z = Q[5];
		3'b110: Z = Q[6];
		3'b111: Z = Q[7];
	endcase
end
    \end{verbatim}
}
\textit{In these always blocks, `Q` is the current state of the shift register, `nextState` is the next state of the shift register, `S` is the data input for the shift register, and `Z` is the output of the multiplexer. The `enable` signal is used to control whether the shift register should shift the data or not. The `clk` signal is the clock signal for the shift register. The `A`, `B`, and `C` inputs are used to select which output of the shift register should be reflected on the output `Z`.\\}

\rule{\linewidth}{0.4pt}
\caption{Example of the \textit{always blocks} for task \textit{ece241\_2013\_q12}.}
\label{merge1}

\end{figure}

\begin{figure}[!h]

\rule{\linewidth}{0.4pt}
\fontsize{10}{12pt}\selectfont

\textbf{Merged codes:}

{
\fontsize{7pt}{5pt}\selectfont
\begin{verbatim}
module top_module (
	input clk,
	input enable,
	input S,
	input A,
	input B,
	input C,
	output reg Z
);
	// Declare the shift register
	reg [7:0] Q;

	// Declare the next state of the shift register
	reg [7:0] nextState;

	// Update the state of the shift register
	always @(posedge clk) begin
		if(enable) begin
			Q <= nextState;
		end
	end

	// Determine the next state of the shift register
	always @(Q, S) begin
		nextState = {Q[6:0], S};
	end

	// Determine the output Z
	always @(Q, A, B, C) begin
		case({A, B, C})
			3'b000: Z = Q[0];
			3'b001: Z = Q[1];
			3'b010: Z = Q[2];
			3'b011: Z = Q[3];
			3'b100: Z = Q[4];
			3'b101: Z = Q[5];
			3'b110: Z = Q[6];
			3'b111: Z = Q[7];
		endcase
	end
endmodule
\end{verbatim}
}

\rule{\linewidth}{0.4pt}
\caption{Example of merged code sample for task \textit{ece241\_2013\_q12}.}
\label{merge2}

\end{figure}

\bibliographystyle{IEEEtran}
\bibliography{bibfile}

\begin{thebibliography}{10}
\providecommand{\url}[1]{#1}
\csname url@samestyle\endcsname
\providecommand{\newblock}{\relax}
\providecommand{\bibinfo}[2]{#2}
\providecommand{\BIBentrySTDinterwordspacing}{\spaceskip=0pt\relax}
\providecommand{\BIBentryALTinterwordstretchfactor}{4}
\providecommand{\BIBentryALTinterwordspacing}{\spaceskip=\fontdimen2\font plus
\BIBentryALTinterwordstretchfactor\fontdimen3\font minus
  \fontdimen4\font\relax}
\providecommand{\BIBforeignlanguage}[2]{{%
\expandafter\ifx\csname l@#1\endcsname\relax
\typeout{** WARNING: IEEEtran.bst: No hyphenation pattern has been}%
\typeout{** loaded for the language `#1'. Using the pattern for}%
\typeout{** the default language instead.}%
\else
\language=\csname l@#1\endcsname
\fi
#2}}
\providecommand{\BIBdecl}{\relax}
\BIBdecl

\bibitem{translation}
B.~Zhang, B.~Haddow, and A.~Birch, ``Prompting large language model for machine
  translation: A case study,'' in \emph{International Conference on Machine
  Learning (ICML)}, 2023, pp. 41\,092--41\,110.

\bibitem{robot}
J.~Wu, R.~Antonova, A.~Kan, M.~Lepert, A.~Zeng, S.~Song, J.~Bohg,
  S.~Rusinkiewicz, and T.~Funkhouser, ``Tidybot: Personalized robot assistance
  with large language models,'' \emph{Autonomous Robots}, vol.~47, no.~8, pp.
  1087--1102, 2023.

\bibitem{codegen}
S.~I. Ross, F.~Martinez, S.~Houde, M.~Muller, and J.~D. Weisz, ``The
  programmer's assistant: Conversational interaction with a large language
  model for software development,'' in \emph{International Conference on
  Intelligent User Interfaces (IUI)}, 2023, pp. 491--514.

\bibitem{survey}
T.~Chen, G.~L. Zhang, B.~Yu, B.~Li, and U.~Schlichtmann, ``Machine learning in
  advanced ic design: A methodological survey,'' \emph{IEEE Design \& Test},
  vol.~40, pp. 17--33, 2023.

\bibitem{benchmark}
S.~Thakur, B.~Ahmad, Z.~Fan, H.~Pearce, B.~Tan, R.~Karri, B.~Dolan-Gavitt, and
  S.~Garg, ``Benchmarking large language models for automated {V}erilog {RTL}
  code generation,'' in \emph{Design, Automation and Test in Europe Conference
  and Exhibition (DATE)}, 2023, pp. 1--6.

\bibitem{humanaid1}
J.~Blocklove, S.~Garg, R.~Karri, and H.~Pearce, ``Chip-chat: Challenges and
  opportunities in conversational hardware design,'' in \emph{ACM/IEEE Workshop
  on Machine Learning for CAD (MLCAD)}, 2023, pp. 1--6.

\bibitem{trained2}
M.~Liu, T.-D. Ene, R.~Kirby, C.~Cheng, N.~Pinckney, R.~Liang, J.~Alben,
  H.~Anand, S.~Banerjee, I.~Bayraktaroglu \emph{et~al.}, ``Chip{N}e{M}o:
  Domain-adapted llms for chip design,'' \emph{arXiv preprint
  arXiv:2311.00176}, 2023.

\bibitem{trained1}
S.~Liu, W.~Fang, Y.~Lu, Q.~Zhang, H.~Zhang, and Z.~Xie, ``{RTL}coder:
  Outperforming gpt-3.5 in design {RTL} generation with our open-source dataset
  and lightweight solution,'' \emph{arXiv preprint arXiv:2312.08617}, 2023.

\bibitem{trained3}
S.~Thakur, B.~Ahmad, H.~Pearce, B.~Tan, B.~Dolan-Gavitt, R.~Karri, and S.~Garg,
  ``Veri{G}en: A large language model for {V}erilog code generation,''
  \emph{ACM Transactions on Design Automation of Electronic Systems}, vol.~29,
  no.~3, pp. 1--31, 2024.

\bibitem{trained4}
Z.~Pei, H.-L. Zhen, M.~Yuan, Y.~Huang, and B.~Yu, ``Better{V}: Controlled
  verilog generation with discriminative guidance,'' \emph{arXiv preprint
  arXiv:2402.03375}, 2024.

\bibitem{trained5}
K.~Chang, K.~Wang, N.~Yang, Y.~Wang, D.~Jin, W.~Zhu, Z.~Chen, C.~Li, H.~Yan,
  Y.~Zhou \emph{et~al.}, ``Data is all you need: Finetuning {LLM}s for chip
  design via an automated design-data augmentation framework,'' \emph{arXiv
  preprint arXiv:2403.11202}, 2024.

\bibitem{gpt4}
J.~Achiam, S.~Adler, S.~Agarwal, L.~Ahmad, I.~Akkaya, F.~L. Aleman, D.~Almeida,
  J.~Altenschmidt, S.~Altman, S.~Anadkat \emph{et~al.}, ``{GPT-4} technical
  report,'' \emph{arXiv preprint arXiv:2303.08774}, 2023.

\bibitem{incontext}
S.~Min, X.~Lyu, A.~Holtzman, M.~Artetxe, M.~Lewis, H.~Hajishirzi, and
  L.~Zettlemoyer, ``Rethinking the role of demonstrations: What makes
  in-context learning work?'' in \emph{Conference on Empirical Methods in
  Natural Language Processing (EMNLP)}, 2022, pp. 11\,048--11\,064.

\bibitem{rag}
Y.~Mao, P.~He, X.~Liu, Y.~Shen, J.~Gao, J.~Han, and W.~Chen,
  ``Generation-augmented retrieval for open-domain question answering,'' in
  \emph{International Joint Conference on Natural Language Processing
  (IJCNLP)}, 2021, pp. 4089--4100.

\bibitem{raghdl1}
Y.~Fu, Y.~Zhang, Z.~Yu, S.~Li, Z.~Ye, C.~Li, C.~Wan, and Y.~C. Lin,
  ``{GPT4AIGChip}: Towards next-generation ai accelerator design automation via
  large language models,'' in \emph{IEEE/ACM International Conference on
  Computer Aided Design (ICCAD)}, 2023, pp. 1--9.

\bibitem{raghdl2}
Y.~Tsai, M.~Liu, and H.~Ren, ``{RTLfixer}: Automatically fixing {RTL} syntax
  errors with large language models,'' \emph{arXiv preprint arXiv:2311.16543},
  2023.

\bibitem{raghdl3}
X.~Yao, H.~Li, T.~H. Chan, W.~Xiao, M.~Yuan, Y.~Huang, L.~Chen, and B.~Yu,
  ``{HDL}debugger: Streamlining {HDL} debugging with large language models,''
  \emph{arXiv preprint arXiv:2403.11671}, 2024.

\bibitem{autochip}
S.~Thakur, J.~Blocklove, H.~Pearce, B.~Tan, S.~Garg, and R.~Karri, ``Autochip:
  Automating {HDL} generation using {LLM} feedback,'' \emph{arXiv preprint
  arXiv:2311.04887}, 2023.

\bibitem{veval}
M.~Liu, N.~Pinckney, B.~Khailany, and H.~Ren, ``Verilog{E}val: Evaluating large
  language models for {V}erilog code generation,'' in \emph{IEEE/ACM
  International Conference on Computer-Aided Design (ICCAD)}, 2023.

\bibitem{hop}
S.~Yang, E.~Gribovskaya, N.~Kassner, M.~Geva, and S.~Riedel, ``Do large
  language models latently perform multi-hop reasoning?'' \emph{arXiv preprint
  arXiv:2402.16837}, 2024.

\bibitem{cot}
J.~Wei, X.~Wang, D.~Schuurmans, M.~Bosma, B.~Ichter, F.~Xia, E.~Chi, Q.~V. Le,
  and D.~Zhou, ``Chain-of-{T}hought prompting elicits reasoning in large
  language models,'' in \emph{Advances in Neural Information Processing
  Systems}, vol.~35, 2022, pp. 24\,824--24\,837.

\bibitem{beamsearch}
A.~Graves, ``Sequence transduction with recurrent neural networks,''
  \emph{arXiv preprint arXiv:1211.3711}, 2012.

\bibitem{gpt35}
``{GPT}-3.5 turbo: Language models,'' Accessed: 2022 [Online]. Available:
  \url{https://platform.openai.com/docs/models/gpt-3-5-turbo}.

\bibitem{iverilog}
S.~Williams and M.~Baxter, ``Icarus {V}erilog: open-source {V}erilog more than
  a year later,'' \emph{Linux Journal}, vol. 2002, no.~99, p.~3, 2002.

\bibitem{passat}
M.~Chen, J.~Tworek, H.~Jun, Q.~Yuan \emph{et~al.}, ``Evaluating large language
  models trained on code,'' \emph{arXiv preprint arXiv:2107.03374}, 2021.

\end{thebibliography}

\end{document}